\documentclass{llncs}
\usepackage{alltt}
\title{Extending the Calculus of Constructions with Tarski's fix-point theorem}
\author{Yves Bertot}
\institute{INRIA Sophia Antipolis}
\newcommand{\coqand}{/\char'134}
\begin{document}
\maketitle

\begin{abstract}
We propose to use Tarski's least fixpoint theorem as a basis to define
recursive functions in the calculus of inductive constructions.  This
widens the class of functions that can be modeled in type-theory based
theorem proving tool to potentially non-terminating functions.  This
is only possible if we extend the logical framework by adding the axioms
that correspond to classical logic.  We claim that the extended
framework makes it possible to reason about terminating and non-terminating
computations and we show that common facilities of the calculus of inductive
construction, like program extraction can be extended to also handle the
new functions.
\end{abstract}
\section{Introduction}
For theoretical computer scientists, Tarski's least 
fix-point theorem is a simple basic block to
assert the existence of objects defined by recursive equations.  These
objects may be inductive types and recursive functions
\cite{Huet87,HarrisonInductive}.
However, to use
this theorem, one needs to express that the domain of interest indeed has
the required completeness property and that the function being considered
indeed is continuous.  If the goal is to define a partial recursive function,
then this requires using axioms of classical logic, and for this reason
the step is seldom taken in the user community of type-theory based theorem
proving.    However, the constructive prejudice is not a necessity: adding
classical logic axioms to the constructive logic that is naturally provided
by type theory can often be done safely, in a way that makes it possible
to retain the consistency of the whole system.

In this paper, we suggest working in the setting of classical logic to
increase our capability to reason about potentially non-terminating
recursive functions.  No inconsistency is introduced in the process,
because potentially non-terminating functions of type \(A\rightarrow
B\) are actually modeled as functions of type \(A\rightarrow B_\bot\):
the fact that a function may not terminate is recorded in its type,
non-terminating computations are given the value \(\bot\) which is
distinguished from all the regular values, and one can reason classically
about the fact that a function terminates or not.  This is obviously
non-constructive but does not introduce any inconsistency.

One of the advantages of type-theory based theorem proving is that actual
programs can be derived from formal models, with guarantees that these
programs satisfy properties that are predicted in formally verified proofs.
This derivation process, known as extraction 
\cite{PaulinWerner93,Letouzey2002types},
performs a cleaning operation
so that all parts of the formal models that correspond to compile-time
verifications are removed.  Thanks to this cleaning operation the
extracted programs may actually be reasonably efficient.

In the absence of classical logic axioms, type theory already makes it possible
to model potentially non-terminating functions as total terminating functions
with an extra argument, where this extra argument explicitly states that
the input actually belongs to the function's domain of definition
\cite{bove:tutorial}.
In the
formal model of an algorithm, a potentially non-terminating function can thus
only be invoked if one proves that the particular input for this invocation
is indeed in the domain definition.  Once extracted in a conventional
programming language, this ensures that extracted pieces of software using
these partial functions do use them only when termination is ensured.  It
is still possible to use the extracted functions outside their domain of
definition by deliberately calling them from the toplevel, but inner uses
of the function are guaranteed to terminate.

When axioms are added to the logical framework, three cases may occur:
first, the new axioms may make the system inconsistent; second, the new
axioms may be used only in the part of the models that will be cleaned away
by the extraction process; third, the axioms may be used in the part of
the models that is extracted in the derived programs.  We don't really
need to discuss the first case that should be avoided at all costs.
In the second case, the extraction process still produces a
consistent program, with the same guarantee of termination and this guarantee
relies on reasoning steps that belong to classical logic, which is acceptable
as long as the whole framework remains consistent.  In the third case,
the added axiom needs to be linked to a piece of software that implements
the behavior predicted by the axiom.  We claim that this can be done
safely if Tarski's least fix-point theorem is added as an axiom to the
basic calculus of inductive constructions, along with other axioms to
describe classical logic.

Tarski's least fix-point theorem
can be used to justify the existence of recursive
functions, because these functions can be described as the least fix-point
of the functional\footnote{We consistently use the word {\em functional}
as a noun to describe a higher-order function.} that arises in their
recursive equation.  However, it is
necessary to ensure that the function space has the properties of a
complete partial order and that the functional is continuous.
  These facts can be motivated using a simple development of basic
domain theory.  With the help of a variant of the axiom of choice, this
theorem can be used to produce a function, which we shall call
{\tt Tarski\_fix}, that takes as argument a
functional and a proof that this functional is continuous and returns
a recursive function, which is the least fix-point of this functional.
This recursive function can then be combined with other algorithms
to build larger software models.

With respect to extraction, we propose a program written in the
target functional language that cannot be described in the constructive
part of the calculus of inductive constructions and can be used
as the target code for every use of the function {\tt Tarski\_fix}.
We also suggest a few improvements to the extraction process that should
help make sure that fairly efficient code can be obtained automatically
from the formal models studied inside our extension of the calculus
of inductive constructions.

From the formal proof point of view, Tarski's least fix-point theorem provides
two important properties for the function it produces.  The first property
is that the result function satisfies the fix-point equation used to
define it.  This fix-point equation is important to express how the
computation evolves in one step of recursion.  We already advocated the
importance of fix-point equations in previous work
\cite{BalaaBertot02}.  This fix-point equation is useful when
we want to prove that under some conditions a function is guaranteed to
terminate.  The second property is
that the result function is the least fix-point of the functional of interest.
As a corollary, it is also the least upper bound of a sequence of functions
that is obtained by iterating the functional on the bottom element of the
complete partial order \(A\rightarrow B_\bot\).  As such, it benefits
from tools that make it possible to reason by induction on the
length of computations, thus providing what is called \emph{fix-point}
induction in \cite{Winskel93}.  This makes it possible to prove properties
of the result of functions when it exists, it can also be used to prove
that under some conditions a function will fail to terminate.

In this paper, we recapitulate an easy proof of Tarski's least fix-point
theorem.  We show how the conditions of applicability of this theorem can
be proved for the formal description of potentially non-terminating
recursive functions, we then describe an small example.
  In particular, this example contain
proofs about recursive functions as a support to discuss the techniques
that are available.  The next section discusses matters related to extraction
and execution of the recursive programs that are thus obtained.  Related
work and opportunities for further extensions are reviewed at the end
of the paper.  All the experiments described in this paper were done
with coq \cite{coq-8.0,BertotCasteran04} and can be found
on the internet from the author's web page\footnote{The current address
prefix is {\tt http://www-sop.inria.fr/marelle/Yves.Bertot}, the proofs are
available through the file {\tt proofs.html}.}.

\section{Proving the fix-point theorem}
The statement of the theorem we are interested in is the following:
\begin{theorem}
  In a complete partial order with a minimal element, every continuous
function has a fix-point.
\end{theorem}
This is the form that is found in courses on programming language semantics
like \cite{NielsonNielson92,Winskel93}.  We
formalized the easy proof that is found in \cite{NielsonNielson92},
in the calculus of inductive constructions without axioms.  Thus, this
theorem is part of constructive mathematics when we are in a type
that can constructively be described as a complete partial order.

To make sense from this theorem's statement, we need to describe the
meaning of the various concepts.  A {\em complete} partial order is a partial
order (a type with a binary relation that is reflexive, antisymmetric,
and transitive), with the extra property that every {\em chain}
 has a least upper
bound, where a chain is a sequence \(u_n\)
such that \(\forall n, R~u_n~u_{n+1}\); it is a trivial matter to show
that the least upper bound of a sequence, when it exists, is unique.  A
{\em continuous} function is a function that preserves least upper bounds:
continuous functions are always considered between two complete partial
orders \((A,R)\) and \((B,R')\).  The function \(f\) is called
continuous if for every chain \(u_n\) in \((A,R)\), when 
\(u\) is its least upper bound, then \(f(u)\) is the least upper
bound of  the sequence \(f(u_n)\).  This definition does
not require that the function \(f\) should be monotonic, but we proved
that continuous functions are necessarily monotonic.

We actually proved the theorem in a context with a collection of local variables
and hypotheses.  These are summarized as: \(A\) is a Type, 
\(R\) is a binary relation, that is reflexive and antisymmetric (it
is actually not required to be transitive) and complete, there is
a minimal element \(\bot\), and \(f\) is continous.
The statement we prove is the following one:
\begin{alltt}
{\sf Theorem} Tarski_least_fixpoint : \(\exists \phi: A, least_fixpoint R f \phi\).
\end{alltt}
The proof is decomposed in five lemmas.  First, we show that the
sequence \(u_n = f^n\bot\) is a chain (by induction on \(n\)), then
we show that any upper bound \(u\) of \(u_n\) is also an upper bound
of \(f(u_n)\) (just by translating indices), and that \(f(u)\) is
an upper bound of \(u_n\), (using the monotonicity of \(f\) and
again by translating indices).  Then we show that the least upper bound
 \(\phi\) of \(u_n\) is a fix-point of \(f\) (since \(f(\phi)\)
is also an upper bound we have \(R~u_n~f(u_n)\), by continuity \(f(\phi)\)
is a least upper bound of \(f(u_n)\), but \(\phi\) is also an upper
bound of \(f(u_n)\), and we conclude by antisymmetry).  Then we show that
any fix-point of \(f\) is necessarily an upper bound of \(u_n\) (by
induction on \(n\)) and this is enough to conclude that the \(\phi\) is
the least fix-point of \(f\).

An important corollary of this proof is that the least fix-point of
\(f\) is also the least upper bound of the sequence \(f^n\bot\).  This
is an important tool for subsequent proofs: this statement will be useful
for the proofs by {\em fix-point} induction.

Specialists in the theory of the calculus of constructions will have noted
that the existential construct in the theorem's statement is the
statement of the sort {\tt Prop}, so the function whose existence is
asserted cannot be used to define other functions of sort {\tt Set} or
{\tt Type}.  To avoid this limitation, we can use a \(\Sigma\)-type
version of the axiom of definite description to transform this
existential statement into a \(\Sigma\)-type (based on the {\tt sigT}
contructor).  This axiom is considered acceptable by specialists, but
one should be careful as it is incompatible with the variant of
the calculus of inductive constructions where the {\tt Set} sort is
impredicative \cite{chicli03quotients}.
Fortunately, this impredicative variant
is not the default form provided in a prover like Coq and most work
is usually done in a predicative setting.

\section{Using the theorem to define recursive functions}
To define recursive functions of type \(A\rightarrow B\), we need to find
a binary relation \(\subseteq\) such that \((A\rightarrow B, \subseteq)\)
really is a complete partial order, and the definition of the recursive
function can be understood as a fix-point equation.  The solution that is
used in domain theory is to add artificially a minimal element to the
type \(B\) and to choose an order structure so that the minimal element
is minimal but any other elements are incomparable.  In conventional
programming languages, this is done easily using a unary {\tt option} type
constructor.  In this paper, we will re-use this {\tt option} type,
with a constructor named {\tt Some}, meant to carry a value of the original
type and a constructor named {\tt None}, which we
shall use as the minimal element, sometimes noted \(\bot\).  In our 
mathematical notations, we shall write \(B_\bot\) for {\tt option \(B\)}.
Thus, {\tt option} constructs a family of types parameterized by the initial
type \(B\).
We then define a family of relations (also parameterized by the type \(B\)).
This relation is named {\tt option\_cpo} in our experiment and will be
noted \(\subseteq\) in this paper (hiding the parameter \(B\) when there
is no ambiguity).  The relation is then defined as the reflexive closure of
the relation such that \(\bot \subseteq x\) always holds.  This family
of relations is easily proved to be a family of complete partial orders.

For any type \(B\), when \(R\) is a binary relation on \(B\), we can {\em lift}
the relation to any function space \(A\rightarrow B\).  This is simply
done by {\em pointwise} transposition of the relation on \(B\): \(f\) is
in relation with \(g\) if \(R(f(x),g(x))\) holds for every \(x\) in \(A\).
We will write \(R_{A\rightarrow B}(f,g)\) or simply \(R(f,g)\) when this is not
ambiguous.  If \(R\) is a complete partial order, it is easy to prove
that \(R_{A\rightarrow B}\) is reflexive and transitive, even in a purely
constructive setting.  However, to prove
that \(R_{A\rightarrow B}\) is antisymmetric, we need an axiom
of extensionality, which is usually not provided in a constructive
logical framework.  This axiom simply states that two functions that are
pointwise equal will be considered equal.   This axiom is quite tame
and we use it in
our experiment without further question.

The next question is to show that lifting preserves completeness.
 Actually, when \(f_n\) is a chain of functions
for \(R_{A\rightarrow B}\), the values \(f_n(x)\) when \(x\) is fixed 
constitue a chain in \(B\).  Since \((B,R)\) is complete, this chain has a
least upper bound and we can thus define a unique value \(v_x\) for every
\(x\) which is the least upper bound of the chain \(f_n(x)\).  This
should be enough to define a new function \(f\) that is the least upper bound
of the chain \(f_n\).  However, going from the existence of \(v_x\) for
every \(x\) to a function of \(x\) requires an extra axiom known as
the axiom of definite description.  It is considered to be a variant of
the axiom of choice (or rather the axiom of unique choice).  This axiom
is also obviously non constructive.  Still, we assume that this axiom
can also be added to our logical framework without endangering the
consistency of the system.

The axiom of definite description has the following statement:
{\em for every relation on \(A\times B\), if for every \(x\) in \(A\), there
is a unique \(y\) in \(B\) such that \(R(x,y)\) holds, then there exists
a function \(f\) such that for every \(x\) \(R(x,f(x))\) holds}.
To compute the least fix-point of a chain of functions \(f_n\), we simply
choose \(R(x,y)\) to mean {\em \(y\) is the least upper bound of the
chain \(f_n(x)\)}.

In general, we want to model a recursive function \(f\) of type 
\(A\rightarrow B_\bot\) and we have to construct a continuous functional
\(F\) 
of type \((A\rightarrow B_\bot)\rightarrow(A\rightarrow B_\bot)\) so
that \(f=F f\).  From now on, we will write \(f\subseteq g\) 
to mean the lifted order from the natural order on \(B_\bot\).  In our formal
development this order will be called {\tt f\_order} and we prove
that it is a complete partial order.  We then specialize the
least fix-point theorem to obtain a function called {\tt Tarski\_fix}
with the following type:\label{function-space}
\begin{alltt}
Tarski_fix:
  forall (A B:Set)(f:(A->option B)->A->option B)
   (Hct : continuous (f_order A B)(f_order A B) f), 
   A->option B.
\end{alltt}
The first two arguments are usually not written when this function
is used.

This function has a companion theorem to express that the function
that is built is the least fix-point of the functional:
\begin{alltt}
Theorem Tarski_fix_prop :
    forall (A B:Set) (f:(A -> option B)->A -> option B)
      (Hct: continuous (f_order A B)(f_order A B) f),
       least_fixpoint (f_order A B) f (Tarski_fix f Hct).
\end{alltt}

The next problem is to show that the functionals we encounter are
continuous.
In practice, our recursive functions will respect a smooth regularity:
the value \(\bot\) is added to the result type to represent the fact that the
function may not terminate.  The condition of continuity on the functional
\(F\) corresponds to this interpretation of 
``potential non termination'': every expression containing a potentially
non-terminating computation should fail to terminate if it actually
uses the value returned by this computation and that computation fails
to terminate.  To use the value of a potentially non-terminating
computation one needs to write a pattern-matching construct on this
computation: the continuity condition will be satisfied if we ensure
that the \(\bot\) value is returned in the \(\bot\) case of this
matching construct.

For example, if we want to define a factorial function we will construct
a functional {\tt F\_fact} as follows:
\begin{alltt}
Definition F_fact (fact:Z -> option Z) (x:Z) : option Z :=
  match Zeq x 0 with
    true => Some 1
  | false => 
    match fact(x - 1) with 
      None => None
    | Some v => Some (x*v)
    end
  end.
\end{alltt}
The proof that this function is continuous is given in a theorem
with the following statement.
\begin{alltt}
Theorem F_fact_continuous : 
   continuous (f_order Z Z)(f_order Z Z) F_fact.
\end{alltt}

Once the continuity proof is completed, we can define the function
with a command of the following form:
\begin{alltt}
Definition fact : Z -> option Z :=
  Tarski_fix F_fact F_fact_continuous.
\end{alltt}

Proving that functionals are continuous can be cumbersome, but
they can usually  be understood as the composite of
elementary continuous functions, and composition can be shown to
preserve continuity.  The study of fragments of functional programming
languages that we want to model makes it possible to isolate the various
constructs that are used in all programs and generic proofs or tactics
can be provided for all these constructs.

For instance we can define a continuous test function for tests
on the type {\tt bool}:
\begin{alltt}
Definition cond (A :Set)
  (t:option bool)(v1 v2:option A) : option A :=
  match t with
    None => None
  | Some true => v1
  | Some false => v2
  end.
\end{alltt}
and we can prove once and for all that this function will preserve
the continuity of the function it combines, with a theorem of the
following form:
\begin{alltt}
Theorem cond_continuous :
  forall A B C t F G,
    continuous (f_order' A B) (f_order' A bool) t ->
    continuous (f_order' A B) (f_order' A C) F ->
    continuous (f_order' A B) (f_order' A C) G ->
    continuous (f_order' A B)(f_order' A C) 
       (fun f x => cond (t f x) (F f x) (G f x)).
\end{alltt}
We believe that the same work can be done systematically for the pattern
matching constructs that are associated with any other basic recursive type.

We can also define a function {\tt Apply} that mimicks the application
of a potentially non-terminating function to a value, also computed
by a potentially non-terminating function.  This is a possible
definition:

\begin{alltt}
Definition Apply (A B:Set)(f: option (A -> option B))
    (v:option A) : option B :=
  match v with
    Some x => match f with Some f' => f' x | None => None end
  | None => None
  end.
\end{alltt}
This function can also be provided with a continuity statement, which
we do not give here.

With these basic units, the code for our factorial example
is convertible to the following one:
\begin{alltt}
Definition F_fact2 : (Z -> option Z) -> Z -> option Z :=
   fun f z =>
     cond (Some (Zeq_bool z 0))
        (Some 1)
        (Apply (Some (fun v => Some (z*v))) (f (z - 1))).
\end{alltt}
and the proof that the function is continuous boils down to
a traversal of the basic elements that are combined in the function,
it only requires a few lines in our experiments.

Please note that the definition of the {\tt Apply} function actually
makes it precise and that we envision to compute with our potentially
non-terminating functions in a ``call-by-value'' fashion, since the
application of a function to an argument will fail if the computation
of the argument is non-terminating.  A different approach will be needed
to model the execution of programs in a call-by-name strategy.  However,
we feel it is quite satisfactory that we can describe precisely when
a program will fail to terminate for a given execution strategy.

It is not the purpose of this paper to discuss all the work that needs
to be done to provide a usable package to reason about the continuity
of functions. 
 This was already done, for instance in
 \cite{Paulson:LCF,Regensburger95,MuellerNvOS99} and their work can
probably be re-used directly since
we are now working in a classical framework, where the differences between
HOL, Isabelle and type-theory based proof tools are less important.  Still,
we would like to avoid reconstructing data-types from first principles
of domain theory as it is done in \cite{MuellerNvOS99}.  Although this
work is impressive, we would like a smooth integration of the total
functions already available in type theory with the potentially non-terminating
functions.

\section{Proving properties of functions}
With the help Tarski's least fix-point theorem we can now model a larger
collection of functions.  The new functions always return their result
in an option type, what other authors called a pointed complete partial
order or pcpo \cite{MuellerNvOS99}.
We can actually express that some function
fails to terminate, simply by saying that the value it returns is the
minimal element in the option type.  We can also reason about the domain
of definition and show that the function does return a regular value
when the argument lies in a given domain, described by a predicate.  For
instance, we can follow Bove's proposal to describe the function's domain
as an inductive predicate \cite{bove:tutorial}.

One key point in the definition of recursive functions is that they are
presented as the {\em least fix-point} of the functional of interest.
If we are considering a recursive function \(f\) defined from a functional
\(F\), 
this least fix-point is also the least upper bound of the sequence
\(u_n = F^n(\lambda x.\bot)\), and because the order in the target type
is simplistic, we can prove that for every input of the recursive function
such that \(f x\neq \bot\), there exists a value \(a\) and a number \(n\)
such that \(F_n\lambda x.\bot=a\).  This expresses that the value is indeed
computed in a finite number of iterations.  The theorem has the following
statement (in this statement {\tt iter} is the function that computes
\(F^n(a)\) given the type on which \(F\) operates, \(F\) itself, the
number \(n\), and the initial value \(a\).
\begin{alltt}
Theorem Tarski_fix_iterates_witness :
   forall (A B:Set) (f : (A -> option B)->A -> option B)
      (Hct: continuous (f_order A B)(f_order A B) f)(x:A)(v:B),
      Tarski_fix f Hct x = Some v ->
    exists n, iter (A->option B) f n (fun a => None) x = Some v.
\end{alltt}
This theorem thus makes it possible to grab an inductive piece of data
that {\em measures} the computation of the recursive function on some
input when this computation terminates.  This makes proofs by induction
possible.

In one of our experiments we have defined the semantics of a small
programming language in the spirit of \cite{NielsonNielson92}.  We were
able to use Tarski's fix-point theorem to describe the semantics of while
loops as suggested in the book.  We were then able to prove that when
a value is returned, the same computation can be modeled by a natural
semantics derivation, using an encoding of the natural semantics based
on an inductive predicate.  The experiment was also conducted in
\cite{Nipkow98} with similar achievements.

We can also reason on non-terminating computations.  The {\tt witness}
theorem above has a simple corollary for reasoning on non-termination:
\begin{alltt}
Theorem iterates_none_imp_fix_none :
  forall A B f (Hct: continuous (f_order A B)(f_order A B) f) x,
  (forall n, iter _ f n (fun z => None) x = None) ->
  Tarski_fix f Hct x = None.
\end{alltt}
The two theorems together emulate the concept of fix-point induction
as found in previous work.

For instance,
to prove that our function {\tt fact} will never terminate on negative
inputs:
\begin{alltt}
forall n, forall x, x < 0 ->
    iter _ F_fact n (fun z => None) x = None
\end{alltt}
This proof is done by induction on \(n\).  Note that the quantification
over \(x\) is part of the statement that is proved by induction, this
is necessary because \(x\) changes as the computation progresses, while
remaining negative.  This proof takes only a dozen steps.

Using the corollary on non-termination we can conclude with the following
theorem:
\begin{alltt}
Theorem fact_neg_none : forall x, x < 0 -> fact x = None.
\end{alltt}

Even though the machinery of the calculus of construction only accepts
to compute with functions that are constructively guaranteed to terminate,
we can still use the internal reduction mechanism to compute the value
of our potentially non-terminating recursive function on well chose
arguments.  To achieve this trick, we simply need to use approximates
computed with the help of the {\tt iter} function with an arbitary number
of iteration.  If this number is chosen large enough, we may get an
answer of the form {\tt Some v}.  In this case, we know that the value
of the recursive function is an element of the target type {\tt option B}
that is larger than {\tt Some v} for the order we are using.  There are
no other choices than the value {\tt Some v} itself, so that we are
actually guaranteed to have computed the right value.  Of course,
we may have chosen a value that is not large enough.  In this case
the value returned will be {\tt None} and we cannot conclude.  We don't
know whether the recursive function actually diverges for this particular
input or whether a larger number of iterations would have sufficed.

Still this kind of computation is very useful and may be used in 
reflexive tactics for example: it may be that to proof a certain fact
we need to compute with a recursive function obtained through
the least fix-point theorem.  We can do so with a fixed number of
iterations: if a regular value is returned, it is used in the
proof attempt, if a \(\bot\) value is returned, this means the proof
attempt fails to produce a result in the allocated time.  So be it.

\section{Extraction towards functional programming languages}
One of the fine points of our experiment is the fact that we use a
\(\Sigma\)-type form of the definite description axiom.  This means
that the extraction process may encounter uses of this axiom and
will have the problem of finding a relevant piece of target code to
produce for this axiom.

We have a solution to this problem.  We contend that the extraction
mechanism can still behave correctly if the axiom of definite description
is not used elsewhere than in the proof of Tarski's fix-point theorem.
On the other hand, it is better if this theorem itself is considered
as an axiom, because there is a functional value that can represent it
faithfully in conventional functional programming languages, without
endangering the correctness of the extraction mechanism.

More precisely, we think that it is important to prove the Tarski's
least fix-point theorem to make sure it relies only on well understood
axioms; and we did just that in our experiments.  But then, it should
be replaced with a shadow of itself: an axiom with the same type but
no value that the extraction mechanism can use.  Then the extraction
mechanism should be instructed to bind this axiom to a value in
the target functional programming language, which describes exactly the
computation of the fix-point of an arbitrary function with non-termination
if needed.  Here is an example of such a function, written in {\tt Ocaml}
syntax:
\begin{alltt}
let rec tarski_fix f x = f (fun y -> tarski_fix f y) x
\end{alltt}
Please note that this definition clear expresses that the function being
defined is the fix-point of \(f\).  It also clearly expresses that this
is only to be used for recursive functions (not for recursive non-functional
values).
Moreoever, this one-line of code is carefully engineered to avoid falling
in a non-terminating recursive loop, even in a call-by-value setting (as
in {\tt Ocaml}).  In fact, 
\hbox {\tt fun y -> tarski\_fix f y}
 is
a function that is extensionally equal to {\tt tarski\_fix f}, except that
its computation is deferred until it is really given an argument.

In the Coq system, here is how we treat the fix-point theorem: a first
theorem is proved in the traditional setting (even without needing any
classical logic axiom)  This theorem has the following statement:
\begin{alltt}
Tarski_least_fixpoint
     : forall (A : Set) (R : A -> A -> Prop),
       (forall x : A, R x x) -> antisymmetric R -> complete R ->
       forall bot : A, (forall x : A, R bot x) ->
       forall f : A -> A, continuous R R f ->
      exists phi : A, least_fixpoint R f phi
\end{alltt}
This theorem cannot be used to construct values in the {\tt Set} sort, so
that it does not interfere with the extraction mechanism.

We then use the standard collection of classical axioms to prove that
any space of the form \(A\rightarrow_\bot\) forms a complete partial
order.  In particular, we need to have an axiom of extensionality to
conclude on the property of antisymmetry and the axiom of excluded middle
to establish the relation between least upper bounds of chains and
least fix-points.

Once we have the complete partial order structure on function spaces,
we instantiate the least fix-point theorem to this domain and we
shift its existential quantification to a \(\Sigma\)-type, using the
following form of the axiom of description:
\begin{alltt}
Axiom dependent_description' :
    forall (A:Type) (B:A -> Type) (R:forall x:A, B x -> Prop),
      (forall x:A,
       exists y : B x, R x y \coqand{}
           (forall y':B x, R x y' -> y = y')) ->
       sigT (fun f : forall x:A, B x =>
             (forall x:A, R x (f x))).
\end{alltt}
The specialized theorem is called {\tt Tarski\_fix'} and its statement
is the same as the statement of {\tt Tarski\_fix} which we already
described in section~\ref{function-space}.  We prove a theorem
{\tt Tarski\_fix'\_prop}, with the same statement as {\tt Tarski\_fix\_prop}
to establish the relevant properties.

At this point we are satisfied with the proof of the theorem.  We
actually add the values {\tt Tarski\_fix}
and {\tt Tarski\_fix\_prop} as axioms (morally, they are harmless axioms since
we know they can be proved).  From then on, all the
rest of our development only uses these axioms instead of the proofs.

We then instruct the extraction mechanism to map the axiom {\tt Tarski\_fix}
to the well-chosen function.  The directive is as follows.
\begin{alltt}
Extract Constant Tarski_fix => 
   "let rec t f x = f (fun y -> t f y) x in t".
\end{alltt}
From then on, every function defined using {\tt Tarski\_fix} is correctly
extracted to a recursive function.  We claim that this function is guaranteed
to compute as predicted by the models we study in the calculus of
constructions.  In one of our experiments, we described the denotational
semantics of a simple imperative programming language and proved it
sound with respect to an natural semantics specification, in the spirit of
\cite{Nipkow98,Winskel93}.  Once extracted
to ML, this gives a certified interpreter for the language.

There are two improvements that we can propose but for
which we have not been able to produce an a experiment.  The first improvement
is that the fix-point function should actually be inlined directly in
every recursive function that is based on the least fix-point theorem.
In the current situation, the extracted code for our {\tt fact} function
is a direct call to the {\tt tarski\_fix} function with a functional
as first argument.  Because of this every recursive call of the function
we want to model is implemented with two recursive calls in the target
language.  Moreover, this precludes optimizations like tail-recursion
optimisation.

The second improvement concerns useless pattern-matching constructs that
are added in the code to mirror the pattern-matching constructs that
appear in the calculus of constructions models to handle the possibility
that functions may not terminate.  These pattern-matching constructs appear
as matchings on {\tt option} types.  They are useless because no function
ever explicitly produces a {\tt None} value, at least if the models are
written with the discipline that {\tt None} should be used only to
represent failure to terminate and not other causes of failure.  To avoid
these useless pattern-matching constructs, we suggest that a specific type
constructor, other than {\tt option}, should be used to add a bottom
element to types and the use of the bottom constructor from this specific
type should be forbidden outside the bottom clauses of pattern-matching
constructs on the values of types obtained by this new type constructor.
Actually, this discipline can be verified syntactically by the extraction
tool itself.  In this manner, the pattern-matching constructs related
to non-termination would still appear in the models, but they
would be absent in the extracted code.

The intuition behind this improvement is that the bottom value is only
produce at the end of all times by functions that do not terminate.  It
is safe to discard this possibility, because the eventuality of such
a value really arising in a program will only happen at the end of all
times, in other words, never.
\section{Related work}
The work described here contributes to all the work that has been made
to ease the description and formal proofs about general recursive
functions.  A lot of efforts were put in providing relevant collections
of inductive types equiped with terminating computation inspired from
the notions of primitive recursion
\cite{Camilleri-Melham,Moh93,Aczel77}.  In particular,
it was shown that the notion of accessibility or noetherian induction
could be described using an inductive predicate with a single constructor
in \cite{nordstrom88}.  This accessibility predicate makes it possible to
encode well-founded recursion, when one can prove that all elements of
the input type satisfy the accessibility predicate for a well-chosen
relation (such a relation is called well-founded).  In practice, it requires
ingenuity to find the right well-founded relation
for the function being considered.  If it is not true
that all elements are accessible (or if one cannot exhibit a well-founded
relation that suits the function being defined), the recursive function may
still be
defined, but it will have a well-defined values only for the elements
that can be proved to be accessible for some relation.
This idea was further refined in \cite{DuboisDonzeau1998,bove:tutorial},
where termination is not
described using an accessibility predicate, but directly with an inductive
predicate that actually describes exactly those inputs for which the function
terminates.

In previous work \cite{BalaaBertot02,BertotCasteran04},
we attempted to provide tools that stay closer to the level of
expertise of programmers in conventional functional programming.  The
key point is to start from the recursive equation and to generate
the recursive function definition from this equation.  Users still need
to prove that the recursive calls happen on predecessors of the initial
input for a chosen well-founded relation, but these requirements appear
as proof obligations that are generated as a complement of the
recursive equation.  The tool produces the recursive function and
a proof of the recursive equation.  The technique is also based on
iterating the functional that occurs in the recursive equation,
but no continuity argument is required (instead we use
a well-founded relation).  With respect to all
this body of work, our work is original in that it concentrates on
describing potentially non-terminating functions, not by adding extra 
input arguments to describe the domain, but by adding an element to the
target type to denote non-termination.

We have only done the minimal amount of domain theory to just make
it possible to define potentially non-terminating fuctions and
perform basic reasoning steps on these functions.  More complete
studies of domain theory have been performed in the LCF system 
\cite{Paulson:LCF}.  It was
also formalized in Isabelle's HOL instanciation to provide a package
known as HOLCF \cite{Regensburger95,MuellerNvOS99}.  We believe
these other experiments can give us guidelines to make it easier
for programmers to prove the continuity requirements.

In early version of the Calculus of constructions, formalizations
of Tarski's least fixpoint theorem were also used to show how
inductive definitions could be encoded directly in the pure (impredicative)
calculus of constructions \cite{Huet87}.  In this respect, it is also
worthwhile to mention that \cite{HarrisonInductive} shows how this
theorem can be used to give a definitional justification of inductive
type in higher-order logic.

\section{Conclusion}
There is a popular belief that type-theory based proof tools can only
be used to reason on functions that are total and terminating for all
inputs, because termination of reductions is needed to ensure the
consistency of these system.  One of the contribution of this paper is
to fight this belief by providing yet another way to model potentially
non-terminating functions.  To do so we re-use Tarski's least fix-point
theorem, a well-known theorem of domain theory.

In this paper, there are three claims that deserve a closer look.  The first
claim is that users of type-theory based theorem provers
should relax their attachment
to constructive mathematics and accept non-constructive axioms that do
not endanger the consistency of the logical system.  Actually, it is high
time that a comprehensive set of axioms should be designed to
make it possible to mix type theory (with the
advantage that it contains a quite efficient reduction mechanism that
makes it possible to compute directly in the logical framework) and classical
higher-order logic (with the advantage of representing more directly the
usual concepts of mathematics).  We believe our paper is the first to
show the direct link between the non-constructive domain theory for
computable functions and extraction capabilities.

The second claim we make and should be scrutinized is that modeling an
arbitrary recursive function is tractable in the setting we propose.
In particular, we have been very quick
on the problem of proving continuity, but this continuity problem may be
the really difficulty of this approach.
We hope that the existing body of work on formalized 
domain theory can be of use here.

The third claim we make is that Tarski's least fix-point theorem is
faithfully realized by the little functional value that we have described.
We have given little arguments for this claim and we wonder whether a formal
proof of this statement could be made, for instance by relying on a
formal description of the target language's semantics and proving that
computations in this formal setting do terminate when models compute
to a regular value different from \(\bot\).

\section*{Acknowledgments}
Benjamin Werner and Hugo Herbelin played a significant role in understanding
what form of the axioms of classical logic provide safe extensions of the
calculus of constructions.  This work also benefited from early experiments
by Kuntal Das Barman and suggestions by P. Aczel.  The author also wishes
to remember the late Gilles Kahn, who started work on formalizing domain
theory in the context of the calculus of inductive constructions in
1996 \cite{Kahn-geom}.
\bibliography{a}
\bibliographystyle{plain}

\end{document}